\documentclass[aps,prl,twocolumn,superscriptaddress]{revtex4-1}
\usepackage{graphicx}
\usepackage{amssymb}
\usepackage{amsmath}
\usepackage{multirow}
\newcommand*\Bell{\ensuremath{\boldsymbol\ell}}

\begin{document}
\title{Superfluid $^3$He in Planar Aerogel}
\author{V.\,V.\,Dmitriev}
\email{dmitriev@kapitza.ras.ru}
\affiliation{P.L.~Kapitza Institute for Physical Problems of RAS, 119334 Moscow, Russia}
\author{M.\,S.\,Kutuzov}
\affiliation{Metallurg Engineering Ltd., 11415 Tallinn, Estonia}
\author{A.\,Y.\,Mikheev}
\affiliation{ Department of Physiology, Perelman School of Medicine, University of Pennsylvania, PA 19104 Philadelphia, USA}
\author{V.\,N.\,Morozov}
\thanks{Deceased.}
\affiliation{Institute of Theoretical and Experimental Biophysics of RAS, 142290 Puschino, Russia}
\author{A.\,A.\,Soldatov}
\affiliation{P.L.~Kapitza Institute for Physical Problems of RAS, 119334 Moscow, Russia}
\author{A.\,N.\,Yudin}
\affiliation{P.L.~Kapitza Institute for Physical Problems of RAS, 119334 Moscow, Russia}

\date{\today}

\begin{abstract}
We report results of experiments with liquid $^3$He confined in a high porosity anisotropic nanostructure which we call planar aerogel. This aerogel consists of nanofibers (with diameters $\sim10$\,nm) which are randomly oriented in the plane normal to the specific axis. We used two samples of planar aerogel prepared using different techniques. We have found that on cooling from the normal phase of $^3$He the superfluid transition in both samples occurs into an equal spin pairing superfluid phase. NMR properties of this phase qualitatively agree with the properties of the superfluid A phase in the anisotropic Larkin-Imry-Ma state. We have observed differences between results obtained in the presence and absence of solid paramagnetic $^3$He on the aerogel strands. We propose that these differences may be due, at least in part, to a magnetic scattering channel which appears in the presence of solid paramagnetic $^3$He.
\end{abstract}

\maketitle

\section{Introduction}
Liquid $^3$He at ultralow temperatures is an ideal model object for studies of influence of impurities on unconventional superfluidity and superconductivity with $p$-wave, spin-triplet pairing: its Fermi surface is an ideal sphere, its  superfluid phases (A, A$_1$ and B) are well studied, and the superfluid coherence length $\xi_0$ can be varied in range of 20-80\,nm by changing pressure \cite{VW}. Superfluid $^3$He is intrinsically pure but impurities can be introduced into it as a high porosity nanostructures like aerogels. Superfluidity of $^3$He in aerogels observed, for the first time, in silica aerogel with porosity of 98\% \cite{parp95,halp95} which is a fractal structure consisting of of SiO$_2$ strands with diameters of $\approx3$\,nm and an average separation of $\sim100$\,nm. In silica aerogels the superfluid transition temperature of $^3$He ($T_{ca}$) is significantly lower than that in bulk $^3$He ($T_c$) and the observed A-like and B-like superfluid phases have the same order parameters as A and B phases correspondingly \cite{bark00,dmit02,kun07,bun08,dmit10,halp11,halp12,li14,halp13,halp14,halp15}.
It has also been found that the weak global anisotropy of silica aerogels plays an important role in the stability of the observed phases \cite{halp08,zim13}. This anisotropy is created either by a deformation of initially isotropic samples, or during the growth of the aerogel, and results in anisotropy of the mean free path of the $^3$He quasiparticles.

The other class of aerogel-like materials called nematic aerogels \cite{asad15}
exhibits the highest possible degree of the anisotropy \cite{askh12,meln15}: strands of such aerogels are nearly parallel to one another that corresponds to infinite stretching of initially globally isotropic structure consisting of randomly oriented straight strands. In the presence of these highly ordered strands, new superfluid phases of $^3$He are realized -- polar, polar-distorted A and polar-distorted B phases \cite{dmit15,dmit12,dmit14,askh15} -- as it is expected from theoretical works \cite{AI,saul13,fom14,ik15}. It has been found that the boundary conditions for scattering of $^3$He quasiparticles on strands of nematic aerogel are extremely important \cite{dmit18}. In pure $^3$He, the strands are covered with $\sim2$ atomic layers of paramagnetic solid $^3$He \cite{Sch87,Rich,Godf,Coll}. In this case the scattering is expected to be nearly diffusive and spin is not conserved during scattering due to a fast exchange between atoms of liquid and solid $^3$He resulting in a magnetic scattering channel \cite{SS,BK,AI2}. Adding a small amount of $^4$He to the cell replaces solid $^3$He on the strands and changes the boundary conditions: firstly, it excludes the magnetic channel; secondly, the scattering remains diffusive only at $P\gtrsim 25$\,bar, while at low pressures it should be specular or partly specular \cite{Rich,P91,K93}. In experiments described in Ref.~\cite{dmit18} the superfluid transition temperature was significantly lower in pure $^3$He than in the case of $^4$He preplating and the polar phase was observed only in the absence of paramagnetic solid $^3$He on the strands. This may indicate the importance of the magnetic channel, the influence of which on polar and polar-distorted states has been considered in theoretical works \cite{Fom,Min}.

A noticeable influence of the boundary condition on superfluid states of $^3$He has recently been observed also in silica aerogel with fairly strong anisotropy which has the same orienting effect on the order parameter as the nematic aerogel \cite{halp20}. At the same time, experiments in $^3$He in isotropic or weakly anisotropic silica aerogels show no significant influence of the boundary conditions on the superfluid phase diagram. In particular, the observed superfluid phases correspond to the A and B phases of bulk $^3$He, regardless of the presence or absence of $^4$He and only a small change in $T_{ca}$ was detected at low pressures presumably due to change of the scattering specularity \cite{halp96,Golov,d2003}. The above-mentioned observations
raise the question of how the influence of the boundary conditions on superfluid states depends on the anisotropy of the aerogel. To answer this question, we have carried out experiments in $^3$He in a new aerogel-like material which we call a planar aerogel. It consists of the strands uniformly distributed in a plane perpendicular to the specific axis $z$ -- the case opposite to nematic aerogel.

\section{Theoretical predictions}
According to Volovik's model \cite{vol08} the planar aerogel corresponds to an infinite uniaxial squeezing of an initially isotropic aerogel considered as a system of randomly oriented cylindrical strands. In this case the A phase having the order parameter
\begin{equation}
\label{A}
A_{\nu k}=\Delta_0d_\nu\left(m_k+in_k\right),
\end{equation}
is expected to emerge \cite{AI,vol08}. Here ${\bf d}\perp{\bf H}$ is the unit spin vector characterizing nematic ordering in the spin subsystem, $\bf H$ is external steady magnetic field, $\bf m$ and $\bf n$ are mutually orthogonal unit orbital vectors representing the ferromagnetic ordering in the orbital subsystem, and $\Delta_0$ is the superfluid gap parameter. A random force induced by aerogel strands may destroy the long-range order in the orbital space forming the Larkin-Imry-Ma (LIM) state of orbital vector $\Bell={\bf m}\times{\bf n}$ which remains spatially inhomogeneous only at distances of $\xi_{LIM}$ which for silica aerogels is $\lesssim1$\,$\mu$m \cite{dmit10,halp13,vol08}. In an isotropic aerogel $\Bell$ is random at longer distances forming the isotropic LIM state. The uniaxial squeezing along $z$ should modify the chaotic spatial distribution of $\Bell$ and make the LIM state anisotropic:
\begin{equation}
\label{vectorl}
\left<\Bell\right>=0,~\left<\ell_z^2\right>=\frac{1+2q}{3},~\left<\ell_x^2\right>=\left<\ell_y^2\right>=\frac{1-q}{3},
\end{equation}
where $\left<\cdot\right>$ is the space average, $1>q>0$ is the parameter of anisotropy ($q=0$ in isotropic aerogel). For deformations of aerogel greater than some critical value the uniform distribution of $\Bell$ ($q=1$) with $\Bell$ fixed along $z$ is expected to be more favorable \cite{vol08}.

Vector $\bf d$ can be either spatially uniform (spin nematic, SN, state), formed under conventional cooling from the normal $^3$He, or randomized in the plane perpendicular to $\bf H$ (spin glass, SG, state), obtained by cooling through $T_{ca}$ with a sufficiently large resonant radio-frequency excitation \cite{dmit10}. The SN state corresponds to a global energy minimum, while the SG state is metastable.

In NMR experiments the magnetization is homogeneous at distances of a dipole length $\xi_D\sim 10\,\mu m$. If $\xi_{LIM}\ll \xi_D$ then a separate resonance line is observed and an identification of superfluid phases of $^3$He in aerogel can be made by measurements of the NMR frequency shift ($\Delta\omega$) from the Larmor value ($\omega_L$), which for the A phase in the LIM state is
\begin{equation}
\label{SN}
2\omega_L\Delta\omega=q\Omega_A^2\left(-\cos\beta+\sin^2\mu\frac{7\cos\beta+1}{4}\right)
\end{equation}
for the SN state and
\begin{equation}
\label{SG}
2\omega_L\Delta\omega=q\Omega_A^2\cos\beta\left(\frac{3}{2}\sin^2\mu-1\right)
\end{equation}
for the SG state \cite{dmit10}. Here $\Omega_A=\Omega_A(T,P)\propto \Delta_0$ is the Leggett frequency, $\beta$ is the tipping angle of magnetization $\bf M$, $\mu$ is the tilt angle of $\bf H$ from $z$-axis. Eqs.~\eqref{SN} and \eqref{SG} with $q=1$ are also applicable for the state with uniform $\Bell$. In linear continuous wave (cw) NMR $\cos\beta \approx 1$, so
\begin{equation}
\label{SNcw}
2\omega_L\Delta\omega=q\Omega_A^2\left(2\sin^2\mu-1\right)
\end{equation}
for the SN state and
\begin{equation}
\label{SGcw}
2\omega_L\Delta\omega=q\Omega_A^2\left(\frac{3}{2}\sin^2\mu-1\right)
\end{equation}
for the SG state.

Due to suppression of the superfluid transition temperature in aerogel ($\delta T_{ca}=T_c-T_{ca}>0$), $\Omega_A$ is smaller than $\Omega_{A0}$, the Leggett frequency of the A phase in bulk $^3$He which is measured, e.g., in Refs.~\cite{A01,A02}. This can be explained by scattering models which describe impurity effects on superfluid $^3$He in globally isotropic aerogel \cite{HISM,IISM}. One of them, the so-called homogeneous isotropic scattering model (HISM), suggests that the suppression of the order parameter (the gap or the Leggett frequency) scales to the transition temperature:
\begin{equation}
\label{suppr}
\Omega_A(\tau)=\frac{T_{ca}}{T_c}\Omega_{A0}(\tau),
\end{equation}
where $\tau$ is the temperature normalized to corresponding superfluid transition temperatures. Experiments in $^3$He in silica aerogel show that the observed suppression is greater \cite{impeff}. These results are explained by the inhomogeneous isotropic scattering model (IISM), which includes specific aerogel parameters. Both HISM and IISM models are not applicable for planar aerogels, but for lack of the adequate theory we are forced to use Eq.~\eqref{suppr} in order to estimate $\Omega_A$ which should not lead to a big error under condition $\delta T_{ca}\ll T_c$.

The A phase belongs to the class of equal spin pairing (ESP) phases whose susceptibilities equal the normal phase value. In the B phase the susceptibility is lower, making it simple to distinguish between these two groups of phases. According to Ref.~\cite{AI}, in the case of a planar aerogel the superfluid transition into the A phase is expected at all pressures and then, on further cooling, into the B phase. In this paper properties of the B phase are not studied in detail, and we focus on investigation of ESP phases in planar aerogels.

\section{Samples and methods}
We have used two samples of planar aerogel prepared by different techniques.

\begin{figure}[t]
\centerline{\includegraphics[width=\columnwidth]{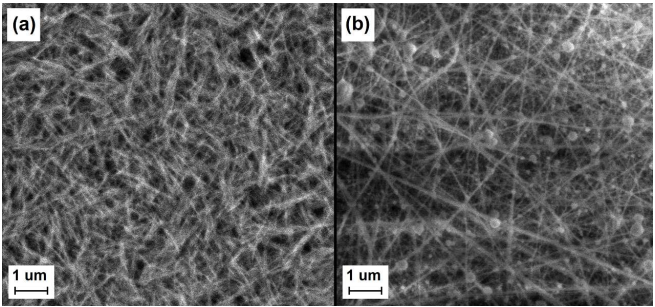}}
\caption{Scanning electron microscope images of the free surface of mullite (a) and nylon (b) planar aerogel samples.}
\label{sem_push}
\end{figure}

The first sample was produced from an aluminum silicate (mullite) nematic aerogel consisting of strands with diameter of $\sim10$\,nm [Fig.~\ref{sem_push}a]. It is a fibrous network in the plane with porosity 88\%, overall density 350\,mg/cm$^3$, and with characteristic lengths of separate strands of $\sim1$\,$\mu$m. Spin diffusion measurements at 2.9\,bar in normal $^3$He confined by this sample in presence of $^4$He on the strands confirm its strong anisotropy: the ratio of spin diffusion coefficients parallel and perpendicular to the specific plane of planar aerogel in the zero temperature limit was found to be 1.64, which is close to the theoretically predicted value of 1.97 \cite{planar_diff}. Effective mean free paths of $^3$He quasiparticles along and normal to the plane of the aerogel are found to be 116\,nm and 71\,nm. The particular sample used here was a stack of three plates each with thickness of 1\,mm and sizes $4\times4$\,mm.

The second sample of planar aerogel was made from a so-called free-standing nanomat (nanofilter) of nylon as a sandwich of 1000 layers with overall thickness of $\approx150$\,$\mu$m and sizes $\approx4\times4$\,mm. Synthesis of such nanofilters is based on electrospinning technique and described in Ref.~\cite{mikh16}. This aerogel mostly consists of very long strands with diameters 10--20\,nm almost parallel to the specific plane [Fig.~\ref{sem_push}(b)], however, it includes thick strands and macroscopic droplets of nylon (both with diameter $\gtrsim100$\,nm) which do not act as impurities in superfluid $^3$He, but increase the durability of this nanofilter. The presence of these macroscopic objects in the planar aerogel complicates direct measurements of its ``effective'' density and porosity. According to our estimations, the porosity of the sample is at least 99\%.

Samples of mullite and nylon aerogels were placed freely in the separate cells of our experimental chamber with corresponding filling factors of $\sim85$\% and $\sim60$\%. The experimental chamber was made from Stycast-1266 epoxy resin and was similar to that described in Ref.~\cite{dmit12}.

Experiments were carried out using linear continuous wave and pulsed NMR in magnetic fields 3.2--57.9\,mT (corresponding NMR frequencies are 104--1877\,kHz) at pressures 2.9--29.3\,bar. In cw NMR experiments the observed NMR lines from superfluid $^3$He in aerogel and from bulk superfluid $^3$He had distinct frequency shifts and were easily distinguishable. We were able to rotate $\bf H$ by any predefined angle $\mu$. Additional gradient coils were used to compensate the magnetic field inhomogeneity. The necessary temperatures were obtained by a nuclear demagnetization cryostat and measured by a quartz tuning fork. Below $T_c$ the fork was calibrated by Leggett frequency measurements in bulk B phase of superfluid $^3$He. In presence of the bulk A phase the temperature was determined from the NMR shift of the bulk signal.

In our experiments we either used pure $^3$He or had $^4$He coverage on the aerogel strands. In the first case we observed a strong paramagnetic NMR signal at low temperatures, indicating the presence of solid paramagnetic $^3$He on the strands. In experiments with $^4$He coverage we added 1.55\,mmole of $^4$He into the empty experimental chamber at $T\leq100$\,mK and then filled it with $^3$He. This amount of $^4$He was found to be sufficient to avoid the solid $^3$He on the strands at all pressures. According to our estimations it corresponds to $\gtrsim2.5$ atomic layers of $^4$He coverage.

When solid $^3$He covers the aerogel strands, a single NMR resonance is observed, as the NMR frequency becomes a weighted average of NMR frequencies of liquid and solid $^3$He due to the fast exchange mechanism \cite{Coll,Fr}. The magnetization of paramagnetic $^3$He follows the Curie-Weiss law, and at $T\sim T_c$ its magnetization ($M_s$) may exceed that of liquid $^3$He ($M_l$), affecting measurements of the NMR frequency shift ($\Delta\omega$) in liquid $^3$He. Moreover, due to the global anisotropy of planar aerogel the frequency shift in solid $^3$He ($\Delta\omega_s$) due to demagnetizing field may become significant \cite{kittel}. Fortunately, the shift in superfluid $^3$He is inversely proportional to $H$ while $\Delta\omega_s \propto H$. Therefore, in  experiments with pure $^3$He we mostly used relatively low magnetic fields, where $\Delta\omega_s$ can be neglected. Correspondingly, in order to obtain ``real'' value of $\Delta\omega$ we recalculated the measured frequency shift ($\Delta\omega^\prime$) using the following equation:
\begin{equation}
\label{recalc}
\Delta\omega\approx\Delta\omega^\prime\left(1+\frac{M_s}{M_l}\right),
\end{equation}
where $M_s/M_l$ was determined from measurements of temperature dependence of the intensity of cw NMR absorption line.

\section{Results with mullite sample}
\begin{figure}[t]
\centerline{\includegraphics[width=\columnwidth]{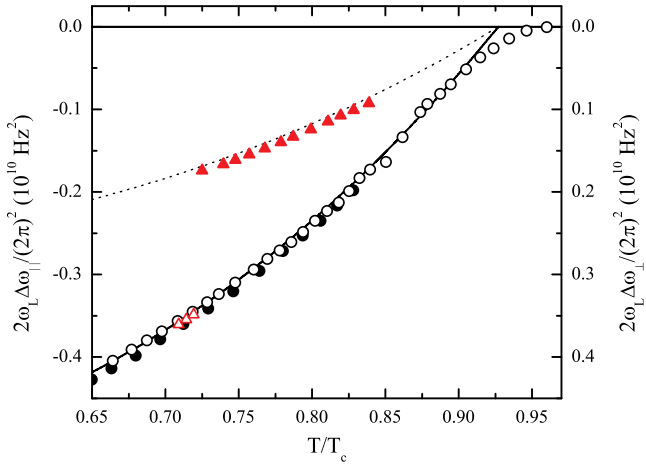}}
\caption{Cw NMR frequency shifts versus temperature in $^3$He in mullite planar aerogel in parallel ($\Delta\omega_\parallel$, circles) and in transverse ($\Delta\omega_\perp$, triangles) magnetic fields in SN (open symbols) and SG (filled symbols) states for the case of $^4$He coverage. Solid and dotted lines are theoretical predictions according to Eqs.~\eqref{SNcw} and \eqref{SGcw} with $q=0.6$, where $\Omega_A$ is calculated from Eq.~\eqref{suppr} using the known temperature dependence of $\Omega_{A0}$ with $T_{ca}=0.927\,T_c$. Triangles and filled circles: $\omega_L/(2\pi)=453$\,kHz. Open circles: $\omega_L/(2\pi)=1303$\,kHz. $P=29.3$\,bar. The $x$-axis represents the temperature normalized to the superfluid transition temperature of bulk $^3$He ($T_c$). Note the different scales on the left and right $y$-axes.}
\label{mullite_SNSG}
\end{figure}
\begin{figure}[t]
\centerline{\includegraphics[width=0.95\columnwidth]{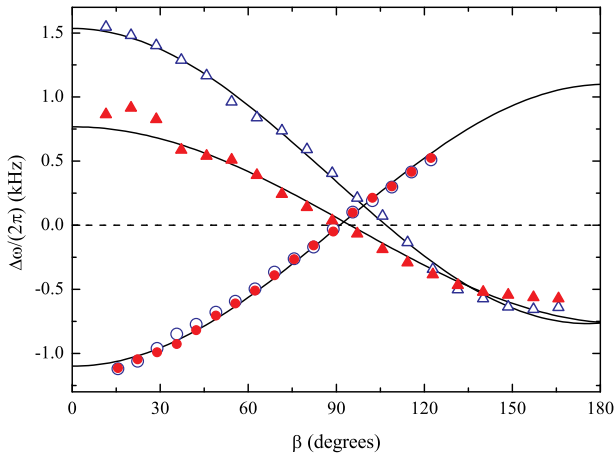}}
\caption{Pulsed NMR frequency shifts versus $\beta$ in $^3$He in mullite planar aerogel for the case of $^4$He coverage in parallel ($\mu=0$, $T=0.88T_c$, circles) and in transverse ($\mu=\pi/2$, $T=0.86T_c$, triangles) magnetic fields in SN (open symbols) and SG (filled symbols) states. $\omega_L/(2\pi)=453$\,kHz. Solid lines correspond to Eqs.~\eqref{SN} and \eqref{SG} with $q\Omega_A^2$ obtained from cw NMR measurements. $P=29.3$\,bar.}
\label{mullite_angle}
\end{figure}
\begin{figure}[t]
\centerline{\includegraphics[width=\columnwidth]{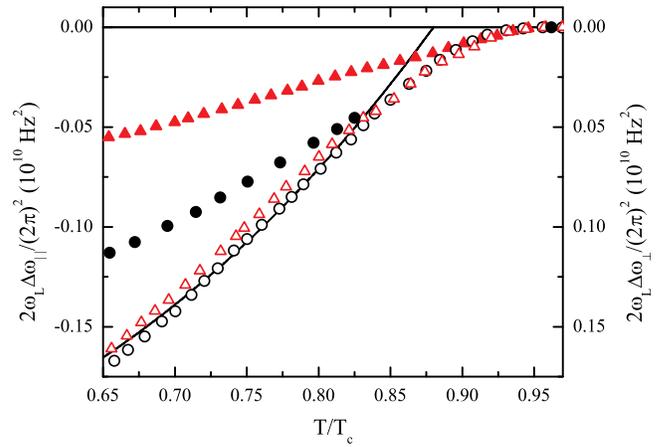}}
\caption{Cw NMR frequency shifts versus temperature in pure $^3$He in mullite planar aerogel in parallel ($\Delta\omega_\parallel$, circles) and in transverse ($\Delta\omega_\perp$, triangles) magnetic fields in SN (open symbols) and SG (filled symbols) states. Solid line is the theoretical prediction for SN states according to Eqs.~\eqref{SNcw} and \eqref{suppr} with $q=0.287$ and $T_{ca}=0.88\,T_c$. $P=29.3$\,bar. $\omega_L/(2\pi)=242$\,kHz.}
\label{mullite3He}
\end{figure}
\begin{figure}[t]
\centerline{\includegraphics[width=\columnwidth]{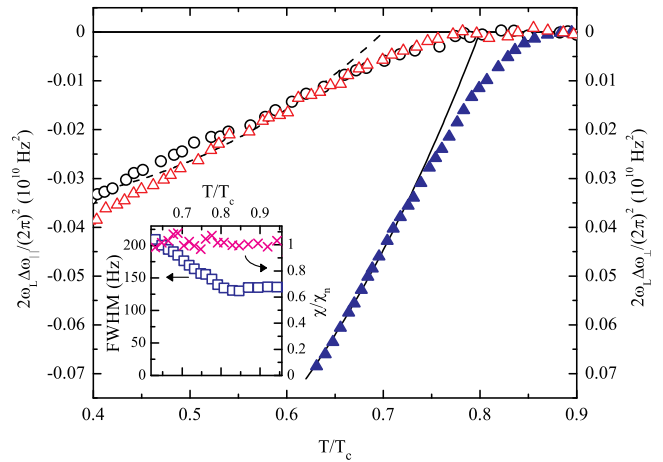}}
\caption{Cw NMR frequency shifts versus temperature in $^3$He confined by mullite planar aerogel in parallel ($\Delta\omega_\parallel$, circles) and in transverse ($\Delta\omega_\perp$, triangles) magnetic fields in SN states for $^4$He coverage ($\omega_L/(2\pi)=453$\,kHz, filled triangles) and pure $^3$He ($\omega_L/(2\pi)=104$\,kHz, open symbols). Lines are theoretical predictions according to Eq.~\eqref{SNcw} and Eq.~\eqref{suppr} with $T_{ca}=0.80\,T_c$, $q=0.46$ (solid line) and $T_{ca}=0.70\,T_c$, $q=0.19$ (dashed line).
Inset: temperature dependencies of the full width at half maximum (FWHM, squares) of cw NMR absorption line and spin susceptibility normalized to the normal state value in $^3$He (crosses) corresponding to the filled triangles in the main graph. $P=7.1$\,bar.}
\label{mullite_SN_lowP}
\end{figure}
\begin{figure}[t]
\centerline{\includegraphics[width=\columnwidth]{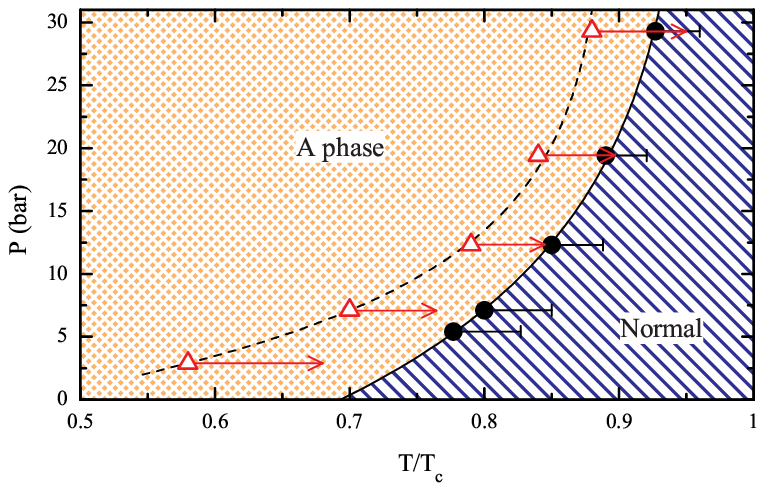}}
\caption{Phase diagram of $^3$He in mullite planar aerogel. Circles mark $T_{ca}$ for $^4$He coverage. Triangles mark $T_{ca}$ for pure $^3$He. Right ends of the error bars mark $T_{ca}^*$. The A phase persists down to the lowest attained temperatures ($\approx0.3T_c$). Solid and dashed lines are guides to eye.}
\label{mullite_PD}
\end{figure}
On cooling from the normal phase of $^3$He in mullite planar aerogel we observe the superfluid transition accompanied by an appearance of nonzero $\Delta\omega$. The transition occurs to the ESP phase, because the magnetic susceptibility of this phase equals the normal phase value and does not depend on $T$. We assume that the observed ESP phase is the A phase in the anisotropic 3D LIM state with the average orientation of the vector $\Bell$ along $z$.
In Fig.~\ref{mullite_SNSG} by open and filled symbols we show temperature dependencies of cw NMR frequency shifts measured with $^4$He coverage at two orientations of $\bf H$ ($\Delta\omega_\parallel$ at $\mu=0$ and $\Delta\omega_\perp$ at $\mu=\pi/2$). Open symbols correspond to cooling through $T_{ca}$ with low NMR excitation, while filled symbols represent measurements obtained after cooling through $T_{ca}$ with high NMR excitation. It is seen that the temperature width of the superfluid transition is rather large ($\sim0.03\,T_c$). We think that this is due to local inhomogeneities of the sample with a characteristic length less than $\xi_D$. The superfluid transition in $^3$He inside regions with higher porosity occurs at higher temperatures than in the main part of the sample and at $T=T_{ca}^*$ a small NMR frequency shift appears which does not follow Eq.~\eqref{suppr}. In order to compare the results with theoretical expectations, we fit the data with Eqs.~\eqref{SNcw}, \eqref{SGcw} and \eqref{suppr} only well below $T_{ca}^*$ (at $T<0.9\,T_{ca}^*$) using $T_{ca}$ and $q$ as fitting parameters. In doing this we obtain good agreement with the theory: (i) $\Delta\omega (\tau) \propto \Omega_{A0}^2 (\tau)$. (ii) In case of cooling through $T_{ca}$ with low excitation we get the SN state and $\Delta\omega_\parallel=-\Delta\omega_\perp<0$ as it follows from Eq.~\eqref{SNcw}. (iii) In case of cooling through $T_{ca}$ with high excitation we get the SG state with $\Delta\omega_{\parallel(SG)}=\Delta\omega_{\parallel(SN)}$ and $\Delta\omega_{\perp(SG)}\approx-0.5\Delta\omega_{\parallel(SN)}$ in accordance with Eqs.~\eqref{SNcw} and \eqref{SGcw}. (iv) Results of pulsed NMR experiments are well described by Eq.~\eqref{SN} (see Fig.~\ref{mullite_angle}).

In pure $^3$He the width of the superfluid transition is larger than in the case of $^4$He coverage (Fig.~\ref{mullite3He}) but the low temperature data also agree with the assumption that we obtain the A phase in the anisotropic LIM state with the average orientation of the vector $\Bell$ along $z$:
in the SN state $\Delta\omega_\parallel=-\Delta\omega_\perp<0$ (open circles and open triangles) and in the SG state $\Delta\omega_{\perp}\approx-0.5\Delta\omega_{\parallel}$ (filled circles and filled triangles) as follows from
Eqs.~\eqref{SNcw} and \eqref{SGcw}. The discrepancy with the theory is that absolute values of $\Delta\omega$ in the SG states are $\approx30$\% smaller than is expected from the data in the corresponding SN states (according Eqs.~\eqref{SNcw} and \eqref{SGcw} open and filled circles in Fig.~\ref{mullite3He} should coincide). The reason for this is not clear, and we can only assume that it is due to an orientational effect of magnetic boundary conditions on vectors $\bf d$ suggested in \cite{AI2}. We also note that in pure $^3$He, in contrast to the case of $^4$He coverage, the SG states are not stable at $T_{ca}\lesssim T<T_{ca}^*$ where they are transformed into the SN states, which is confirmed by measurements of the cw NMR shift in the case of cooling back to low temperatures.

From data shown in Figs.~\ref{mullite_SNSG} and \ref{mullite3He} we obtain that for $^4$He coverage $q=0.6$, $T_{ca}=0.927\,T_c$, and $T_{ca}^*\approx 0.96\,T_c$, while in the case of pure $^3$He $q=0.287$, $T_{ca}=0.88\,T_c$, and $T_{ca}^*\approx 0.95\,T_c$. Thus, the $^4$He coverage increases $q$, $T_{ca}$ and $T_{ca}^*$. Similar behavior was observed at all experimental pressures. At $P=29.3$\,bar the difference in $T_{ca}^*$ between cases of $^4$He coverage and pure $^3$He is very small, but at lower pressures the increase of $T_{ca}^*$ in presence of $^4$He is clearly observable (see Figs.~\ref{mullite_SN_lowP} and \ref{mullite_PD}).

\section{Results with nylon sample}
\begin{figure}[t]
\centerline{\includegraphics[width=\columnwidth]{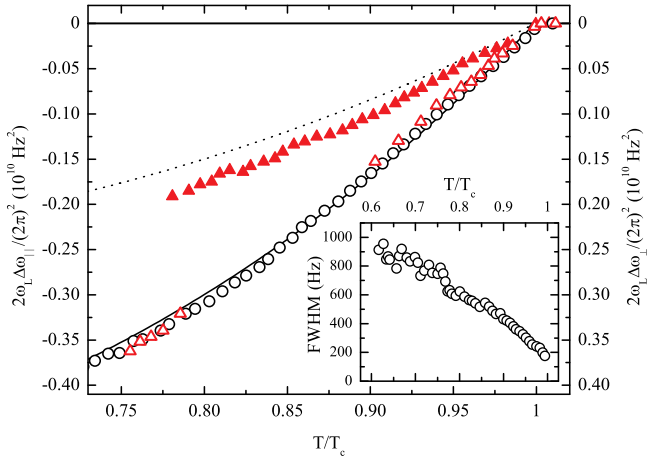}}
\caption{Cw NMR frequency shifts versus temperature in $^3$He confined by nylon planar aerogel in parallel
($\Delta\omega_\parallel$, circles) and in transverse ($\Delta\omega_\perp$, triangles) magnetic fields in SN (open symbols) and SG (filled symbols) states for the case of $^4$He coverage. Circles: $\omega_L/(2\pi)=1269$\,kHz. Triangles: $\omega_L/(2\pi)=460$\,kHz. Lines are theoretical predictions according to Eq.~\eqref{suppr}, Eq.~\eqref{SNcw} (solid line), and Eq.~\eqref{SGcw} (dotted line) with $T_{ca}=0.999\,T_c$ and $q=0.265$. $P=29.3$\,bar. Inset: temperature dependence of FWHM in the cw NMR measurements corresponding to circles in the main graph.}
\label{nylon_SNSG}
\end{figure}

\begin{figure}[t]
\centerline{\includegraphics[width=\columnwidth]{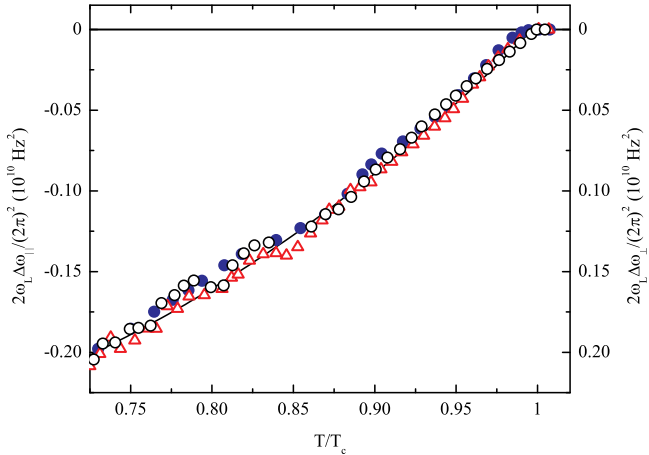}}
\caption{Cw NMR frequency shifts versus temperature in $^3$He confined by nylon planar aerogel in parallel ($\Delta\omega_\parallel$, circles) and in transverse ($\Delta\omega_\perp$, triangles) magnetic fields in the SN state for the case of pure $^3$He. Open circles and triangles: $\omega_L/(2\pi)=460$\,kHz. Filled circles: $\omega_L/(2\pi)=1269$\,kHz. The solid line is a theoretical prediction according to Eq.~\eqref{SNcw} and Eq.~\eqref{suppr} with $T_{ca}=0.997\,T_c$ and $q=0.265$. $P=29.3$\,bar.}
\label{nylon_SN_3he}
\end{figure}

\begin{figure}[t]
\centerline{\includegraphics[width=\columnwidth]{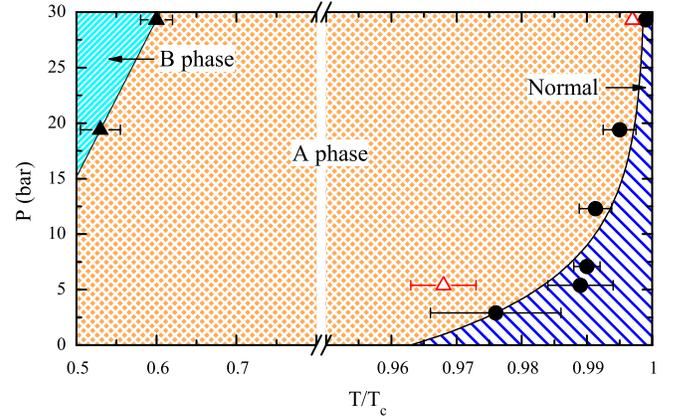}}
\caption{Phase diagram of $^3$He in nylon planar aerogel. Circles and open triangles mark $T_{ca}$ for the cases of $^4$He coverage and pure $^3$He respectively. Filled triangles mark the transition from A phase to B phase on cooling for both cases. Lines are guides to eye.}
\label{nylon_PD}
\end{figure}

At all experimental pressures, upon cooling from the normal phase of $^3$He in nylon planar aerogel the superfluid transition occurs also into the ESP phase. The temperature width of the superfluid transition was found to be very small ($\sim0.001\,T_c$), so for the nylon sample $T_{ca}\approx T_{ca}^*$. In Figs.~\ref{nylon_SNSG} and \ref{nylon_SN_3he} we show results of cw NMR experiments at $P=29.3$\,bar for $^4$He coverage and pure $^3$He respectively. The spin dynamics in the ESP phase is, again, well described by Eqs.~\eqref{SNcw} and \eqref{SGcw} for the A phase having the vector $\Bell$ mostly oriented along $z$: $\Delta\omega_{\parallel(SN)}=-\Delta\omega_{\perp(SN)}<0$ and $\Delta\omega_{\perp(SG)}\approx-0.5\Delta\omega_{\parallel(SN)}$. The suppression of $T_{ca}$ in the nylon sample is small due to its very high porosity (Fig.~\ref{nylon_PD}) and at 29.3\,bar no essential difference in $T_{ca}$ between cases of pure $^3$He and $^4$He coverage is seen, but the value of $q$ in pure $^3$He is significantly smaller. At low pressure the difference in $T_{ca}$ becomes larger, similar to the case of the mullite sample.
\begin{figure}[t]
\centerline{\includegraphics[width=\columnwidth]{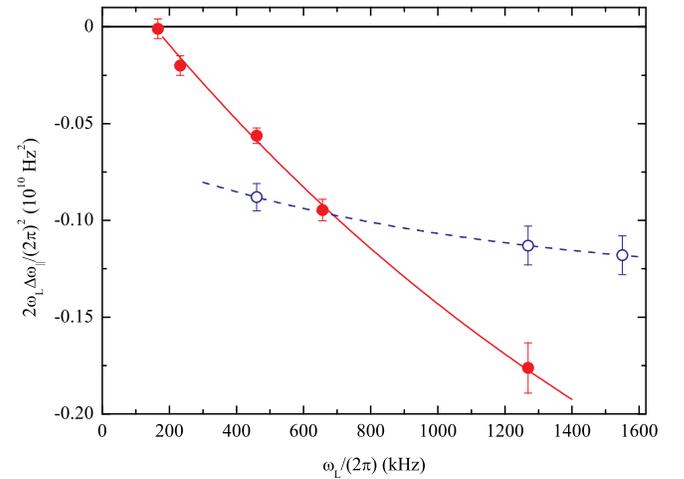}}
\caption{Cw NMR frequency shift of superfluid $^3$He in nylon planar aerogel in parallel magnetic field versus $\omega_L$ for the cases of $^4$He coverage (open circles) and pure $^3$He (filled circles). $T=0.7\,T_c$, $P=5.4$\,bar. Solid and dashed lines are guides to eye.}
\label{nylonH}
\end{figure}
We note two essential differences in comparison with the case of mullite sample.

Firstly, the width of the cw NMR line in the superfluid phase is much larger in comparison with the mullite sample (see the insets in Figs.~\ref{nylon_SNSG} and \ref{mullite_SN_lowP}) indicating that the observed A phase is in the LIM state with  $\xi_{LIM}\gtrsim\xi_D$ due to relatively long characteristic distance between strands ($\xi_a$) because from Ref.~\cite{vol08} it follows that $\xi_{LIM}\propto \xi_0^2 \xi_a$. At low pressures $\xi_{LIM}$ increases due to the increase of $\xi_0$ and in this case the width of the cw NMR line becomes comparable with $\Delta\omega$.

Secondly, at low pressures the effective NMR shift ($2\omega\Delta\omega$) increases with the increase of the magnetic field value (Fig.~\ref{nylonH}). This effect is more prominent at $\mu=0$ and in pure $^3$He.
In the case of $^4$He coverage this effect is noticeably smaller and was not observed at $P>10$\,bar.
In pure $^3$He, experiments were done only at two pressures (5.4 and 29.3\,bar) and at 29.3\,bar the effective shift was also independent on the magnetic field value. The reason of such behavior is not clear and requires further investigations. However, this effect indicates that the value of the field and boundary conditions influence the anisotropy of the LIM state in this sample.

At high pressures ($P\gtrsim15$\,bar) upon cooling in the A phase we observe a rapid drop in intensity of the NMR line, indicating a transition to the B phase, as predicted in Ref.~\cite{AI}. In this paper we do not study this phase in detail.

\section{Discussion}
\begin{figure}[t]
\centerline{\includegraphics[width=\columnwidth]{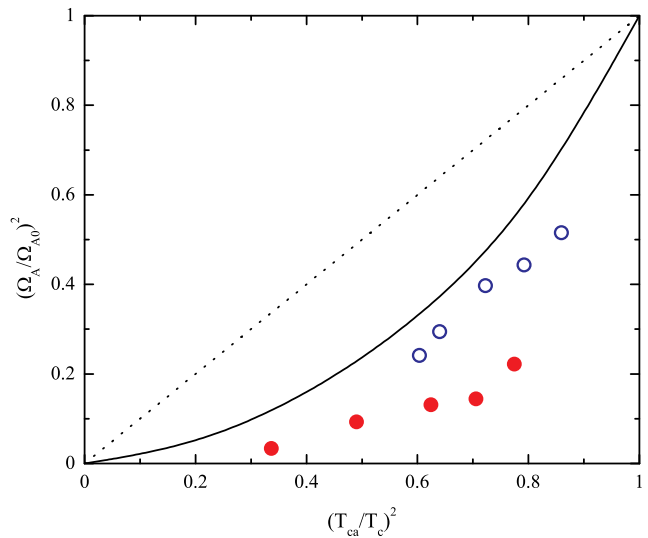}}
\caption{Suppression of the A phase order parameter in mullite planar aerogel in assumption of $q=1$, obtained in cw NMR experiments at different pressures from 2.9 to 29.3\,bar, versus the transition temperature for the cases of full $^4$He coverage (open circles) and pure $^3$He (filled circles), compared to the HISM with unitary scattering shown as a dotted line (Eq.~\eqref{suppr}) and to the IISM plotted as a solid line to match measurements in the A phase in silica aerogel \cite{impeff}.}
\label{mullite_suppr}
\end{figure}

In our experiments we measure $q\Omega_A^2$ and then estimate $q$ using Eq.~\eqref{suppr}. Surely, this is the correct procedure for the nylon sample and the results indicate the formation of the LIM state in this sample. In the mullite sample the condition $\delta T_{ca}\ll T_c$ is not fulfilled, and, in principle, the suppression of $\Omega_A$ may be substantially larger than it follows from Eq.~\eqref{suppr}, as it occurs in isotropic silica aerogel \cite{IISM,impeff}. Therefore, it is possible that $q=1$ and in the mullite sample we obtain the state with the spatially homogeneous $\Bell\parallel z$. In Fig.~\ref{mullite_suppr} we show dependencies of $q\left(\Omega_A/\Omega_{A0}\right)^2$ on $T_{ca}$ obtained from our NMR measurements in $^3$He in the mullite sample. Since $\Omega_A\propto \Delta_0$ then from Fig.~\ref{mullite_suppr} it follows that, in assumption of $q=1$, the suppression of the order parameter is significantly larger than is predicted by the HISM model and also than the suppression in $^3$He in 98\% silica aerogel, which is well described by the IISM model. Therefore, we think it is more likely that in the mullite sample we also obtain the LIM state. Unfortunately, existing theoretical models are not applicable to the anisotropic scattering which occurs in planar aerogel and further development of the theory is necessary for treatment of our results.

In the absence of the magnetic scattering channel, the suppression of the transition temperature and of the gap in $^3$He in planar aerogel should be determined by the effective mean free path of $^3$He quasiparticles in $x-y$ plane ($\lambda_{xx}$) \cite{saul13}, which is decreased with a decrease of the specularity of the quasiparticle scattering \cite{planar_diff}. This may explain the difference in superfluid transition temperatures at low pressures in cases of pure $^3$He (nearly diffusive scattering) and $^4$He coverage (nearly specular scattering). However, at $P\gtrsim 25$\,bar the scattering is expected to be nearly diffusive regardless of the presence or absence of $^4$He coverage. Correspondingly, we propose that at least at high pressures the magnetic scattering is responsible for the suppression of $T_{ca}$ in the mullite sample. The additional suppression of the value of $q\Omega_A^2\propto q\Delta_0^2$ in experiments with pure $^3$He in comparison with the case of $^4$He coverage is observed in both samples, but our NMR measurements do not allow us to distinguish changes in $q$ and in $\Omega_A^2$ separately. Therefore, at least at 29.3\,bar (farthest right experimental points in Fig.~\ref{mullite_suppr}) the observed difference in suppression of $q\Omega_A^2$ may be due to not only the suppression of the order parameter, but also to an influence of the magnetic scattering on the anisotropy of the LIM state.

\section{Conclusions}
We performed NMR experiments in $^3$He confined in two samples of new aerogel-like material called as planar
aerogel. This aerogel has nearly maximal possible global anisotropy, corresponding infinite uniaxial compression of an initially globally isotropic structure consisting of randomly oriented straight strands.
Overall, the experimental data indicate that in both samples the superfluid transition of $^3$He at all pressures occurs into the A phase in the LIM state with a preferred orientation of $\Bell$ along $z$. This result agrees with theoretical predictions \cite{AI,saul13} and with experiments in anisotropic silica aerogel \cite{halp12}, where the region of existence of the A phase was also extended to low pressures in the case of the same orienting effect on $\Bell$. We have also observed differences between results obtained in the presence and absence of solid paramagnetic $^3$He on the aerogel strands, i.e., the additional suppression of cw NMR frequency shift and of $T_{ca}$ in the presence of solid $^3$He. At a pressure of 29.3\,bar these differences cannot be explained by the change in specularity of the $^3$He quasiparticles scattering alone. This fact allows us to assume that magnetic scattering may play an important role here.

It is worth noting that in the planar aerogel the anisotropy of scattering of $^3$He quasiparticles is analogous to that in a narrow gap, and in the recent experiments with $^3$He-A in a very thin gap \cite{saun19} it was suggested that the suppression of the superfluid transition temperature in the presence of solid paramagnetic $^3$He on the walls is partially due to the magnetic scattering.

Thus we conclude that the boundary conditions for scattering of $^3$He quasiparticles essentially influence the superfluid phase diagram of $^3$He in a planar aerogel. As in the case of $^3$He in nematic aerogel, the influence is more noticeable in the case of lower porosity.

\begin{acknowledgments}
The preparation of the mullite sample was made by M.S.K. The synthesis of the nylon sample was made by A.Y.M and V.N.M. The investigation of $^3$He in planar aerogels was carried out by V.V.D., A.A.S., and A.N.Y. and supported by grant of the Russian Science Foundation (project \#\,18-12-00384). We are grateful to I.A.~Fomin, E.V.~Surovtsev and A.~Cember for useful discussions and comments.
\end{acknowledgments}


\begin{thebibliography}{99}
\bibitem{VW} D.~Vollhardt and P.~W$\ddot{o}$lfle, {\it The Superfluid Phases of Helium~3} (Tailor \& Francis, 1990).
\bibitem{parp95} J.V.~Porto and J.M.~Parpia, Phys. Rev. Lett. {\bf 74}, 4667 (1995).
\bibitem{halp95} D.T.~Sprague, T.M.~Haard, J.B.~Kycia, M.R.~Rand, Y.~Lee, P.J.~Hamot, and W.P.~Halperin, Phys. Rev. Lett. {\bf 75}, 661 (1995).
\bibitem{bark00} B.I.~Barker, Y.~Lee, L.~Polukhina, D.D.~Osheroff, L.W.~Hrubesh, and J.F.~Poco, Phys. Rev. Lett. {\bf 85}, 2148 (2000).
\bibitem{dmit02} V.V.~Dmitriev, V.V.~Zavjalov, D.E.~Zmeev, I.V.~Kosarev, and N.~Mulders, JETP Lett. {\bf 76}, 312 (2002).
\bibitem{kun07} T.~Kunimatsu, T.~Sato, K.~Izumina, A.~Matsubara, Y.~Sasaki, M.~Kubota, O.~Ishikawa, T.~Mizusaki, and Yu.M.~Bunkov, JETP Lett. {\bf 86}, 216 (2007).
\bibitem{bun08} J.~Elbs, Yu.M.~Bunkov, E.~Collin, H.~Godfrin, and G.E.~Volovik, Phys. Rev. Lett. {\bf 100}, 215304 (2008).
\bibitem{dmit10} V.V.~Dmitriev, D.A.~Krasnikhin, N.~Mulders, A.A.~Senin, G.E.~Volovik, and A.N.~Yudin, JETP Lett. {\bf 91}, 599 (2010).
\bibitem{halp11}  J.~Pollanen, J.I.A.~Li, C.A.~Collett, W.J.~Gannon, and W.P.~Halperin, Phys. Rev. Lett. {\bf 107}, 195301 (2011).
\bibitem{halp12} J.~Pollanen, J.I.A.~Li, C.A.~Collett, W.J.~Gannon, W.P.~Halperin, and J.A.~Sauls, Nat. Phys. {\bf 8}, 317 (2012).
\bibitem{li14} J.I.A.~Li, A.M.~Zimmerman, J.~Pollanen, C.A.~Collett, W.J.~Gannon, and W.P.~Halperin, J. Low Temp. Phys. {\bf 175}, 31 (2014).
\bibitem{halp13} J.I.A.~Li, J.~Pollanen, A.M.~Zimmerman, C.A.~Collett, W.J.~Gannon, and W.P.~Halperin, Nat. Phys. {\bf 9}, 775 (2013).
\bibitem{halp14} J.I.A.~Li, A.M.~Zimmerman, J.~Pollanen, C.A.~Collett, W.J.~Gannon, and W.P.~Halperin, Phys. Rev. Lett. {\bf 112}, 115303 (2014).
\bibitem{halp15} J.I.A.~Li, A.M.~Zimmerman, J.~Pollanen, C.A.~Collett, and W.P.~Halperin, Phys. Rev. Lett. {\bf 114}, 105302 (2015).
\bibitem{halp08} J.~Pollanen, K.R.~Shirer, S.~Blinstein, J.P.~Davis, H.~Choi, T.M.~Lippman, W.P.~Halperin, and L.B.~Lurio, J. Non-Cryst. Solids {\bf 354}, 4668 (2008).
\bibitem{zim13} A.M.~Zimmerman, M.G.~Specht, D.~Ginzburg, J.~Pollanen, J.I.A.~Li, C.A.~Collett, W.J.~Gannon, and W.P.~Halperin, J. Low Temp. Phys. {\bf 171}, 745 (2013).
\bibitem{asad15}  V.E.~Asadchikov, R.Sh.~Askhadullin, V.V.~Volkov, V.V.~Dmitriev, N.K.~Kitaeva, P.N.~Martynov, A.A.~Osipov, A.A.~Senin, A.A.~Soldatov, D.I.~Chekrygina, and A.N.~Yudin, JETP Lett. {\bf 101}, 556 (2015).
\bibitem{askh12} R.Sh.~Askhadullin, V.V.~Dmitriev, D.A.~Krasnikhin, P.N.~Martynov, L.A.~Melnikovsky, A.A.~Osipov, A.A.~Senin, and A.N.~Yudin, J. Phys. Conf. Ser. {\bf 400}, 012002 (2012).
\bibitem{meln15} V.V.~Dmitriev, L.A.~Melnikovsky, A.A.~Senin, A.A.~Soldatov, and A.N.~Yudin, JETP Lett. {\bf 101}, 808 (2015).
\bibitem{dmit15} V.V.~Dmitriev, A.A.~Senin, A.A.~Soldatov, and A.N.~Yudin, Phys. Rev. Lett. {\bf 115}, 165304 (2015).
\bibitem{dmit12} R.Sh.~Askhadullin, V.V.~Dmitriev, D.A.~Krasnikhin, P.N.~Martynov, A.A.~Osipov, A.A.~Senin, and A.N.~Yudin, JETP Lett. {\bf 95}, 326 (2012).
\bibitem{dmit14} V.V.~Dmitriev, A.A.~Senin, A.A.~Soldatov, E.V.~Surovtsev, and A.N.~Yudin, JETP {\bf 119}, 1088 (2014).
\bibitem{askh15} R.Sh.~Askhadullin, V.V.~Dmitriev, P.N.~Martynov, A.A.~Osipov, A.A.~Senin, and A.N.~Yudin, JETP Lett. {\bf 100}, 662 (2015).
\bibitem{AI} K.~Aoyama and R.~Ikeda, Phys. Rev. B {\bf 73}, 060504(R) (2006).
\bibitem{saul13} J.A.~Sauls, Phys. Rev. B {\bf 88}, 214503 (2013).
\bibitem{fom14} I.A.~Fomin, JETP {\bf 118}, 765 (2014).
\bibitem{ik15} R.~Ikeda, Phys. Rev. B {\bf 91}, 174515 (2015).
\bibitem{dmit18} V.V.~Dmitriev, A.A.~Soldatov, and A.N.~Yudin, Phys. Rev. Lett. {\bf 120}, 075301 (2018).
\bibitem{Sch87} A.~Schuhl, S.~Maegawa, M.W.~Meisel, and M.~Chapellier, Phys. Rev. B {\bf 36}, 6811 (1987).
\bibitem{Rich} M.R.~Freeman and R.C.~Richardson, Phys. Rev. B {\bf 41}, 11011 (1990).
\bibitem{Godf} J.A.~Sauls, Yu.M.~Bunkov, E.~Collin, H.~Godfrin, and P.~Sharma, Phys. Rev. B {\bf 72}, 024507 (2005).
\bibitem{Coll} E.~Collin, S.~Triqueneaux, Yu.M.~Bunkov, and H.~Godfrin, Phys. Rev. B {\bf 80}, 094422 (2009).
\bibitem{SS}J.A.~Sauls and P.~Sharma, Phys. Rev. B {\bf 68}, 224502 (2003).
\bibitem{BK} G.~Baramidze and G.~Kharadze, J. Low Temp. Phys. {\bf 135}, 399 (2004).
\bibitem{AI2} K.~Aoyama and R.~Ikeda, J. Phys. Conf. Ser. {\bf 150}, 032005 (2009).
\bibitem{P91} S.M.~Tholen and J.M.~Parpia, Phys. Rev. B {\bf 47}, 319 (1993).
\bibitem{K93} D.~Kim, M.~Nakagawa, O.~Ishikawa, T.~Hata, T.~Kodama, and H.~Kojima, Phys. Rev. Lett. {\bf 71}, 1581 (1993).
\bibitem{Fom} I.A.~Fomin, JETP {\bf 127}, 933 (2018).
\bibitem{Min} V.P.~Mineev, Phys. Rev. B {\bf 98}, 014501 (2018).
\bibitem{halp20} A.M.~Zimmerman, M.D.~Nguyen, J.W.~Scott, and W.P.~Halperin, Phys. Rev. Lett. {\bf 124}, 025302 (2020).
\bibitem{halp96} D.T.~Sprague, T.M.~Haard, J.B.~Kycia, M.R.~Rand, Y.~Lee, P.J.~Hamot, and W.P.~Halperin, Phys. Rev. Lett. {\bf 77}, 4568 (1996).
\bibitem{Golov} A.~Golov, J.V.~Porto, and J.M.~Parpia, Phys. Rev. Lett. {\bf 80}, 4486 (1998).
\bibitem{d2003} V.V.~Dmitriev, I.V.~Kosarev, N.~Mulders, V.V.~Zavjalov, and D.Ye.~Zmeev, Physica (Amsterdam) {\bf 329B–-333B}, 320 (2003).
\bibitem{vol08} G.E.~Volovik, J. Low Temp. Phys. {\bf 150}, 453 (2008).
\bibitem{A01} A.I.~Ahonen, M.~Krusius, and M.A.~Paalanen, J. Low Temp. Phys. {\bf 25}, 421 (1976).
\bibitem{A02} M.R.~Rand, H.H.~Hensley, J.B.~Kycia, T.M.~Haard, Y.~Lee, P.J.~Hamot, and W.P.~Halperin, Physica (Amsterdam) {\bf 194–-196B}, 805 (1994).
\bibitem{HISM} E.V.~Thuneberg, S.K.~Yip, M.~Fogelstr\"om, and J.A.~Sauls, Phys. Rev. Lett. {\bf 80}, 2861 (1998).
\bibitem{IISM} R.~H\"anninen and E.V.~Thuneberg, Phys. Rev. B {\bf 67}, 214507 (2003).
\bibitem{impeff} W.P.~Halperin, H.~Choi, J.P.~Davis, and J.~Pollanen, J. Phys. Soc. Japan {\bf 77}, 111002 (2008).
\bibitem{planar_diff} V.V.~Dmitriev, M.S.~Kutuzov, L.A.~Melnikovsky, B.D.~Slavov, A.A.~Soldatov, and A.N.~Yudin, arXiv:1810.12710.
\bibitem{mikh16} A.Y.~Mikheev, Y.M.~Shlyapnikov, I.L.~Kanev, A.V.~Avseenko, and V.N.~Morozov, Eur. Polym. J. {\bf 75}, 317 (2016).
\bibitem{Fr} M.R.~Freeman, R.S.~Germain, E.V.~Thuneberg, and R.C.~Richardson, Phys. Rev. Lett. {\bf 60}, 596 (1988).
\bibitem{kittel} V.V.~Dmitriev, M.S.~Kutuzov, A.A.~Soldatov, A.N.~Yudin, JETP Lett. {\bf 108}, 816 (2018).
\bibitem{saun19}  P.J.~Heikkinen, A.~Casey, L.V.~Levitin, X.~Rojas, A.~Vorontsov, P.~Sharma, N.~Zhelev, J.M.~Parpia, J.~Saunders, arXiv:1909.04210.

\end{thebibliography}
\end{document}